# Atomic Scale Visualization of Quantum Interference on a Weyl Semimetal Surface

Hao Zheng,[1,*] Su-Yang Xu,[1,*] Guang Bian,[1,*] Cheng Guo,[2] Guoqing Chang,[3,4] Daniel S. Sanchez,[1] Ilya Belopolski,[1] Chi-Cheng Lee,[3,4] Shin-Ming Huang,[3,4] Xiao Zhang,[2] Raman Sankar,[5] Nasser Alidoust,[1] Tay-Rong Chang,[6,1] Fan Wu,[7] Titus Neupert,[8] Fangcheng Chou,[5] Horng-Tay Jeng,[6,9] Nan Yao,[7] Arun Bansil,[10] Shuang Jia,[2,11] Hsin Lin[3,4] and M. Zahid Hasan[1,†]

1. Laboratory for Topological Quantum Matter and Spectroscopy (B7), Department of Physics, Princeton University, Princeton, New Jersey 08544, USA
2. International Center for Quantum Materials, School of Physics, Peking University, China
3. Centre for Advanced 2D Materials and Graphene Research Centre, National University of Singapore, 6 Science Drive 2, Singapore 117546
4. Department of Physics, National University of Singapore, 2 Science Drive 3, Singapore 117542
5. Center for Condensed Matter Sciences, National Taiwan University, Taipei 10617, Taiwan
6. Department of Physics, National Tsing Hua University, Hsinchu 30013, Taiwan
7. Princeton Institute for the Science and Technology of Materials, Princeton University, 70 Prospect Avenue, Princeton, New Jersey 08540, USA
8. Princeton Center for Theoretical Science, Princeton University, Princeton, New Jersey 08544, USA
9. Institute of Physics, Academia Sinica, Taipei 11529, Taiwan
10. Department of Physics, Northeastern University, Boston, Massachusetts 02115, USA
11. Collaborative Innovation Center of Quantum Matter, Beijing, 100871, China

* These authors contributed equally to this work.
† mzhasan@princeton.edu

**Abstract**

Weyl semimetals may open a new era in condensed matter physics, materials science and nanotech after graphene and topological insulators. We report the first atomic scale view of the surface states of a Weyl semimetal (NbP) using scanning tunneling microscopy/spectroscopy. We observe coherent quantum interference patterns that arise from the scattering of quasiparticles near point defects on the surface. The measurements reveal the surface electronic structure both below and above the chemical potential in both real and reciprocal spaces. Moreover, the interference maps uncover the scattering processes of NbP's exotic surface states. Through comparison between experimental data and theoretical calculations, we further discover that the scattering channels are largely restricted by the orbital and/or spin texture of the surface band. The visualization of the scattering processes can help design novel transport effects and electronics on the topological surface of a Weyl semimetal.

Understanding and utilization of the surface of topological quantum matter lead to new physics and potential applications [1–4]. A Weyl semimetal is a new topological phase of matter that extends the topological classification beyond insulators, exhibits quantum anomalies, possess exotic surface Fermi arc electron states and provides the first ever realization of Weyl fermions in physics [5-11]. The surface of a Weyl semimetal has been predicted to exhibit interesting tunneling and transport properties, leading to potential electronic and spintronic applications [12–14]. However, for a long period of time, experimental work has been held back due to the lack of material realization. The very recent discovery of the Weyl semimetal state this year in the TaAs class of inversion breaking materials, including TaAs, TaP, NbAs and NbP, has received worldwide attention [15–22]. So far, the surface states of Weyl semimetals have only been accessed by photoemission experiments. On the other hand, past research on high Tc superconductors, graphene and topological materials have established scanning tunneling microscopy (STM) as a powerful tool in understanding the electronic properties of quantum materials with simultaneous spatial, energy, and momentum resolution [23-30]. However, an STM study on a Weyl semimetal's surface remains lacking.

In this article, we present the first atomic-scale visualization a Weyl semimetal surface via scanning tunneling microscopy/spectroscopy. Our measurements show that scatterings by point defects induce quantum interferences on the surface of a Weyl semimetal NbP. Such quantum interference data demonstrate a powerful way to map the surface band structure both below and above the Fermi energy simultaneously in real and reciprocal spaces, both of which are beyond the current photoemission measurements. Furthermore, we identify the different scattering channels on the Weyl semimetal NbP surface and discovered some of the them are suppressed due to their orbital and/or spin texture, which is therefore explored here for the first time.

Fig. 1a demonstrates a typical surface morphology of our high quality NbP sample, revealing a large, flat surface ($50 \times 50$ nm$^2$) with very few impurities and defects. The close-up STM image in Fig. 1b shows the atomically ordered lattice with only one native point defect on the top atomic layer in the middle of a $10 \times 10$ nm$^2$ region. The STM topography indicates a square lattice with a lattice constant 3.4 Å which is consistent with the X-ray diffraction determined lattice constant a = 3.3 Å [31, 32]. This proves that the cleaving plane is the (001) surface. Moreover, the point defect manifests itself as a missing atom in the occupied states (Fig.1b) but a cross-shape bright feature in the empty states (Fig. 1a). We speculate that it is a phosphorus (P) vacancy on the surface. The simultaneously obtained differential conductance (d$I$/d$V$) map in Fig.1c shows the local density of states (LDOS) near the single point defect identified in the topography image.

The presence of the defect induces clear modulation of the LDOS in real space, leading to a four-leafs clover shaped pattern. Similar to the clover-like pattern on the $Bi_2Se_3$ surface [33], the feature in Fig.1c presents a fingerprint of the Weyl semimetal NbP(001) surface. Additionally, we also observe another type of defect, which appears as yellow spots and have a size larger than the point defect (Fig. 1a). These are likely to be adatoms or impurity clusters. We take spectroscopic measurement on a defect-free region (Fig. 1d). It shows a finite LDOS at the Fermi level, consistent with the existence of metallic surface states. As displayed in the upper panel of Fig. 1d, the calculated LDOS from niobium (Nb) or P atoms in the topmost layer of NbP(001) have different shapes. The measured LDOS ($dI/dV$) is in agreement with calculations for P termination. As displayed in the structure model in Fig.1e, the naturally cleaved NbP(001) surface is indeed terminated by P atom layer, which is consistent with our measurement.

We present some essential results of the bulk and surface electronic structure of NbP that can help understand our STM measurements. There are 24 Weyl nodes in the first bulk Brillouin zone (BZ). On the (001) surface, they project as 16 projected Weyl nodes (Fig. 1f). We name the 8 projected nodes near the $\bar{X}$ ($\bar{Y}$) point as $W_1$ (the small dots in Fig.1f). They have projected chiral charge of $\pm 1$. We name the other 8 nodes near the midpoint of the $\bar{\Gamma} - \bar{X}$ ($\bar{Y}$) lines, with projected chiral charges of $\pm 2$ as $W_2$ (the large dots in Fig.1f). Fig. 1g shows a calculated Fermi surface of the NbP (001) surface states in momentum space, which is consistent with angle-resolved photoemission electron spectroscopy (ARPES) data [21, 22]. We find three dominant features: a tadpole-shaped feature that runs from each midpoint of the $\bar{\Gamma} - \bar{X}$ ($\bar{Y}$) line toward the $\bar{X}$ ($\bar{Y}$) point, two semi-bowtie-shaped pockets at each $\bar{X}$ point and two semi-elliptical pockets at each $\bar{Y}$ point. We note that the number of Fermi arcs terminating on a projected Weyl node must equal the absolute value of its projected chiral charge. Thus, each projected $W_2$ ($W_1$) node is associated with two (one) Fermi arcs. The double Fermi arc associated with each $W_2$ Weyl node arises from the head of the tadpole (Fig. S1a and c). The single Fermi arc associated with the $W_1$ Weyl node is a very short direct line connecting the nearby pair of nodes, and they are too small to be experimentally resolved (Fig. S1b and d). Importantly, our calculation also reveals the orbital texture of the NbP (001) surface states. The Fermi contours can be decomposed into different states, which originate from the three *p*-orbitals of the P atoms and five *d*-orbitals of the Nb atoms. Since the NbP(001) surface is terminated by P atoms, we restrict our discussion to the *p*-orbital derived surface states. The two tadpole-shaped contours near the midpoint of the $\bar{\Gamma} - \bar{X}$ ($\bar{Y}$) line have $p_y$ ($p_x$) character, while the pockets at the zone boundary are mainly derived from the $p_z$ orbital (Fig. S2).

Fourier-transform (FT) d$I$/d$V$ techniques provide a powerful way to simultaneously study the occupied and unoccupied states of the electronic structure in real and reciprocal (momentum) space [34]. This is not possible by ARPES because photoemission can only probe the occupied states of the electronic structure and lacks spatial resolution. The key is to take advantage of elastic scattering off point defects. On an ideal surface, without any disorder or defects, electronic quasiparticles are Bloch eigenstates characterized by wavevector $\vec{k}$ and energy $\varepsilon$. However, a reasonable amount of disorder consisting of impurities or crystal defects causes elastic scattering, which mixes eigenstates of different $\vec{k}$ at the same $\varepsilon(\vec{k})$. In other words, the presence of a point defect can mix any two states in its vicinity, $\vec{k}_1$ and $\vec{k}_2$, on a constant energy contour, if these two states are coupled by a large matrix element. The mixing of eigenstates with different $\vec{k}$ leads to quantum interference patterns in the vicinity of the defect, which can be uniquely observed in a d$I$/d$V$ map because it measures the LDOS in a spatially-resolved fashion. In addition, by Fourier transforming the real space interference patterns, one can gain insight into the electronic structure in momentum space. Furthermore, scattering matrix elements depend on orbital and spin components of the states at different $\vec{k}$, therefore the quantum interference can also reveal orbital and/or spin characters of the electronic states [25-29].

Since the d$I$/d$V$ map can measure the quasiparticle interference (QPI) patterns in real space and the FT-d$I$/d$V$ map gives rise to the same QPI information but in reciprocal space, we now systematically study QPIs on NbP(001) in both spaces. Fig. 2a shows a d$I$/d$V$ map on a 37 × 37 nm$^2$ terrace at a bias voltage of 80 mV. Remarkably, we observe clear standing waves as spatial modulation of the LDOS around defec on the surface. These "ripples" are related to quantum interference effects that arise from scattering of electron Bloch waves in the presence of crystal defects. The spatial oscillation is clearly seen in the d$I$/d$V$ line-cut shown in Fig. 2b. We perform a FT on these d$I$/d$V$ maps to gain insight into the electronic structure in momentum space (Fig. 2c). The elastic scattering that relates two states, $\vec{k}_1$ and $\vec{k}_2$, leads to the interference pattern in the reciprocal space at $\vec{Q} = (\vec{k}_1 - \vec{k}_2)$. We now show the FT-d$I$/d$V$ map in the reciprocal $Q_x - Q_y$ space (Fig. 2c) and divide the dominant features into three groups (Fig. 2d), according to their distinct line shapes in $Q$-space. (*I*) We see an elliptical contour centered at the origin of the map, [($Q_x$, $Q_y$) = (0, 0)]. The long axis is along the $Q_y$ direction. This feature is repeated at Bragg points ($Q_x$, $Q_y$) = (±2π/a, 0) and ($Q_x$, $Q_y$) = (0, ±2π/a). However, we observe that the ones at ($Q_x$, $Q_y$) = (0, ±2π/a) (noted by the red dotted lines in Fig. 2d) have weaker intensities and are barely visible in Fig. 2c. (*II*) Similarly, we find a bowtie-shaped contour centered at the origin and repeated at Bragg points, whose long axis is along the $Q_x$ direction. The bowtie contours at $(Q_x, Q_y) = (\pm 2\pi/a, 0)$ are much dimmer and are marked as blue dotted lines in Fig. 2d. In addition, we observe weak features at the corners of the map, which suggest that both the elliptical and bowtie-shaped contours are also repeated at the corners, $(Q_x, Q_y) = (\pm 2\pi/a, \pm 2\pi/a)$, indicated by the gray dotted line. (*III*) We observe four contours located within each quadrant of the map. The contours have an approximately rectangular shape. In fact, the vertical edges are nearly straight whereas the horizontal ones are curvy. For each contour, the two edges that are further away from the origin (noted by the green dotted lines) are obviously less visible than the two that are closer to the origin (the solid green lines).

We investigate in turn the contribution of each feature in the real space QPI pattern. In order to do so, we isolate one of the features in the FT-d$I$/d$V$ map (insets of Figs. 2e-g) and perform an inverse FT. This frequency-selected Fourier filtering allows us to visualize the real space interferences that arise from a particular contour in the reciprocal $Q$-space. As shown in Fig. 2e, the elliptical contour leads to the dominant long range standing waves. By contrast, the rectangular contours in Fig. 2g generate an intricate checkerboard knit, where the wavelength of the ripples/interferences is much shorter (atomic scale). We note that by plainly resolving both long wavelength and atomic corrugations in real space, one obtain a rich quantum interference pattern in momentum space. Furthermore, all of the interference patterns in Figs. 2e-g can be identified back in the raw data, real space d$I$/d$V$ map in Figure 2.

Next, we investigate the QPI patterns as a function of energy. Fig.3a shows the real space d$I$/d$V$ maps at different bias voltages. Clear quantum interference is seen at all energies. At a bias voltage of −160 mV, which corresponds to the occupied states at a binding energy of 160 meV below the Fermi level, we observe that the interference pattern has quite short wavelengths and mainly exists in the close vicinity of the defects. The pattern near each defect is nearly $C_4$ symmetric. As one increases the bias voltage, we observe that the ripples propagate further away from the defects. It is important to observe the smooth evolution of the wavelength as a function of the bias voltage as in Fig. 3a, because this shows that our d$I$/d$V$ map is indeed resolving the standing waves of two-dimensional surface quasiparticles, rather than simply imaging the wavefunctions of the defects [35]. In the energy range -100mV to +140mV, the dominant wavelength becomes larger with more positive bias voltage (a larger wavelength in real space corresponds to a smaller wavevector $\vec{Q}$ in reciprocal space), and the pattern becomes increasingly $C_4$ breaking. At high bias voltages for mapping the unoccupied states of the electronic structure, such as +200 mV, the dominant interference discussed above is not observed because its wavelength becomes comparable to the size of the image. Some other weak interference with shorter wavelengths is still observable. We emphasize that the clear visualization of surface standing waves (the explicitly resolved several orders of wave peaks and valleys) is not routinely obtained on the other topological materials, and is not a trivial result as well. Combining with atomic manipulation capable of STM, one can derive quantum phase information of the surface quasiparticle wave function [36]. This may open the possibility to extract topological information contained in the Berry phase of the Weyl fermions.

The observed features in real space have a nice correspondence to the FTs in $Q$-space, shown in Fig. 3b. Specifically, the FT-d$I$/d$V$ map at -200 mV is approximately $C_4$ symmetric. However, at more positive bias voltages, e.g. the one at -40 mV, the FT-d$I$/d$V$ maps clearly break $C_4$ symmetry. The $C_4$ breaking surface state electronic structure of NbP can be understood by considering its crystal structure, where the rotational symmetry is implemented as a screw axis that sends the crystal back into itself after a $C_4$ rotation and a translation by $c/2$ along the rotation axis. Therefore, even though rotation axis is normal to the (001) surface, it still breaks $C_4$ symmetry. Furthermore, we observe that the size of the pockets shrinks as one increases the bias voltage, consistent with the increasing wavelength seen in real space. It proves that the surface bands are form by hole-like quasiparticles.

Our high quality samples, sharp cleaved surfaces and optimally-tuned measurement technique allow us to directly measure both the long-range smooth oscillation and the short-range abrupt atomic corrugation in one d$I$/d$V$ map. As a result, the QPI patterns in reciprocal space show smooth, closed contours. This is in contrast to the QPI patterns reported in many other topological materials, which may only consist of sparse points rather than a line-shaped contour. Our ability to measure a complete QPI contour makes the assignment of the interference features easier and more reliable. We compare the theoretically expected $Q$-space QPI pattern that arises from the surface spectral weight of NbP (001) and our experimental data shown in Figure 4. The agreement proves that our QPI data indeed come from the Weyl semimetal's surface states. Fig. 4a shows the calculated joint density of states (JDOS) of the surface Fermi surface. The JDOS corresponds to an interference pattern that includes all of the possible elastic scattering at the same energy between two states at different crystal momenta, without considering how the scattering cross section may be suppressed by spin and orbital matrix element effects. Fig. 4b shows the calculated QPI pattern that further considers the full matrix element effects [29]. The calculations were carried out for nonmagnetic and fully isotropic scattering center.

In order to uncover the scattering processes on the surface of NbP, we consider the following scattering channels, which are marked by vectors $S_1$, $S_2$, $S_3$, $Q_1$, $Q_2$, and $Q_3$, shown in Fig. 4d. Interestingly, we find that all the scattering vectors that involve the tadpole states ($S_1$-$S_3$) are strongly suppressed. Specifically, the red circle in Fig. 4a highlights the feature that arises from scattering $S_1$. Note that this feature is prominent in the JDOS but missing in both our data (Fig. 4c) and the simulation with considered matrix-element effect (Fig. 4b). The result indicates that the bowtie and tadpole pockets have different orbital and/or spin characters that may be orthogonal to each other. Indeed, it is consistent with our calculations, which shows that the bowtie contours arise from the $p_z$ orbital and the tadpole contours arise from the $p_x$ and $p_y$ orbitals (Fig. S2). Similarly, scattering processes between tadpole pockets such as $S_2$ in Fig. 4d are also suppressed by the orbital textures of the surface states. $S_2$ connects to two tadpole pockets which are separately derived from $p_x$ and $p_y$ orbitals, respectively. Consequently, this scattering is absent in the experimental QPI pattern. The other inter-tadpole scattering $S_3$ is invisible in the QPI presumable due to low spectral weight. The absence of those features in $Q$-space is a direct evidence of the spin-orbital effect on electron scattering and the multi-orbital nature of the novel surface state.

Next, we consider the scatterings arise from the prominent surface state contours, namely the scattering vectors $Q_1$-$Q_3$ (Fig. 4d). The intra-pocket scattering inside of the bowtie-shaped pocket ($Q_1$) in $k$-space gives rise to the almost same bowtie-shaped feature at the origin [$(Q_x, Q_y) = (0, 0)$] of $Q$-space. Similarly, the vector $Q_2$ corresponds to the scattering inside of the elliptical contours centered at one $\bar{Y}$ point in $k$-space. It leads to the elliptical feature at the center of $Q$-space. We can also draw a vector which connects two bowtie-shaped contours (or two elliptical contours) located at different $\bar{X}$ ($\bar{Y}$) points. This scattering process generates the replicas of the bowtie-shaped or elliptical features at the boundaries and the corners in $Q$-space. Finally, the scattering vector $Q_3$ connects a bowtie-shaped contour at a $\bar{X}$ point to an elliptical contour at a $\bar{Y}$ point in $k$-space. This induces the square-shaped features at every quadrant of $Q$-space. The agreement between theory and experiment is quite remarkable considering the rich structure of the Fermi surface and QPI pattern.

Since the QPI data arise from the surface states, we can measure the surface band structure both below and above the Fermi level from our STM data. For example, $Q_1$ connects the two end points of the bowtie on the high symmetry line, $\bar{X} - \bar{M}$, in $k$-space. As a result, there is a simple relationship between the scattering vector $\vec{Q}$ and momentum $\vec{k}$, namely $\vec{k} = \vec{Q}/2$ along $\bar{X} - \bar{M}$ direction. This simple relationship allows us to directly obtain the band structure, $\varepsilon(\vec{k})$ along the $\bar{X} - \bar{M}$ line (if one neglects the spin splitting of the bands which is indeed small in NbP comparing to TaAs). The other intra-pocket scattering vector $Q_2$ can give similar information. The experimental data marked in Fig. 4f are in agreement with the model calculation (solid lines) derived from the band structure as shown in Fig. 4e. We therefore can determine the Fermi velocity to be 2.2 eVÅ and 1.6 eVÅ along $\bar{X} - \bar{M}$ direction and along $\bar{Y} - \bar{M}$, respectively.

In summary, we have for the first time visualized the novel surface of a Weyl semimetal with atomic scale spatial resolution. This has enabled us to resolved clear quantum interference patterns simultaneously in real and momentum spaces. The voltage-dependent real-space imaging allows us to clearly resolve the evolution of the surface standing waves. It opens a way to measure the phase of wavefunction in topological quantum matters. Our QPI results in reciprocal space further identify the orbital-selected scattering process. It helps to deeply understand the surface property on the only available Weyl semimetal class TaAs, TaP, NbAs and NbP and shed light on the future's research on Weyl semimetal based materials and devices.

**Materials and Methods**

Single crystalline NbP samples were grown by standard chemical vapor transport methods. The samples were cleaved at 79K in an ultra-high-vacuum chamber and were immediately transferred *in vacuo* to a Unisoku STM operated at 4.6K and ~$10^{-11}$ mbar. Chemical etched Pt/Ir tips were used. Constant-current mode STM images were obtained by applying a voltage to the sample. A lock-in amplifier was employed to get the first differentiated (d$I$/d$V$) signal with a modulation at 1 to 8 mV and 1kHz. The d$I$/d$V$ maps were measured with STM images, as same method as in Refs [26, 27]. We processed the STM images and d$I$/d$V$ maps in the WsXM software. The FTs were performed directly from the raw data. All color bars in the figures are in linear scale. Surface electronic band structures were obtained by Wannier function based Hamiltonian in a semi-infinite slab using the recursive Green's function method. The surface Green's functions are used to calculate the interference patterns (See supplementary text for more details).

**Acknowledgment**

The STM measurements at Princeton University were supported by the Gordon and Betty Moore Foundations EPiQS Initiative through grant GBMF4547 (Hasan). Theoretical calculations at National University of Singapore were supported by the National Research Foundation, Prime

Minister's Office, Singapore, under its NRF fellowship (NRF Award no. NRF-NRFF2013-03). NbP crystal growth was supported by National Basic Research Program of China (grant nos. 2013CB921901 and 2014CB239302) F.C. acknowledges the support provided by MOST-Taiwan under project no. 102-2119-M-002-004. This work was also supported in part by the National Science Foundation-MRSEC program through the Princeton Center for Complex Materials (DMR-1420541; NY, FW). T.-R. C and H.-T. J. are supported by National Sciene Council, Academia Sinica, and National Tsing Hus University, Taiwan, and also thank NCHC, CINC-NTU, and NCTS, Taiwan for technical support. The work at Northeastern University was supported by the US Department of Energy (DOE), Office of Science, Basic Energy Sciences grant number DE-FG02-07ER46352, and benefited from Northeastern University's Advanced Scientific Computation Center (ASCC) and the NERSC supercomputing center through DOE grant number DE-AC02-05CH11231.

**Competing interests**

The authors declare that they have no competing interests.

**Figure captions**

**Figure 1. Crystalline and electronic properties of the Weyl semimetal NbP(001) surface. a**, Large-scale STM image showing the (001)-surface morphology of the vacuum-cleaved single crystalline NbP. Mapping was performed at 600 mV and 700 pA. On the atomically flat surface, two kinds of typical defects are discerned. Adatoms (or clusters) are shown by yellow spots and native point defects by white crosses. **b**, STM topography and **c**, dI/dV map simultaneously acquired on a point defect (presumably a P vacancy). The measurement setting is -120 mV and 700 pA. **d**, The calculated local density of states from the top P (green) and Nb (yellow) atoms on (001)-surfaces of NbP (upper panel) and a measured d$I$ / d$V$ spectrum (lower panel). The set point of the d$I$/d$V$ is 550mV and 900pA. Direct comparison of the curves confirms our sample cleaved at (001)-P surface. **e**, A schematic of a NbP crystal structure. Nb and P atoms are depicted as yellow and green balls, respectively. **f,** A sketch showing the positons of the projected bulk Weyl nodes in the first surface Brillouin zone (not to scale). Large dots represent the well-separated $W_2$ Weyl nodes, whereas the small dots are the $W_1$ nodes which are close to each other. Colors stand for the topological charge polarities. **g**, Calculated Fermi surface of NbP(001) surface in momentum space. The prominent feature are the two semi-bowtie-shaped contours at $\bar{X}$ points, two semi-ellipses at $\bar{Y}$ points and four tadpole-shaped pockets close to at Γ points.

**Figure 2. Correspondence of real-space modulation to reciprocal-space interference patterns. a**, High-resolution d$I$ / d$V$ map (80 mV, 700 pA) clearly showing standing-wave modulation patterns induced by quasiparticle interference. **b**, A d$I$ /d$V$ profile measured along the dotted line in **a**. A long-range standing wave (wavelength about 10 nm) superimposes with corrugations at atomic level. **c**, The fast Fourier transform (FT) of **a**. **d**, A sketch of the main features in FT. Resolved in the reciprocal space are three groups of interference patterns which are depicted by different colors. *I*, ellipses (blue) appearing in the center and two Bragg points: (2π/a , 0) and (−2π/a , 0). *II*, bow-tie-shaped pockets (red) located at (0, 0) and (0, ±2π/a ) (anther two Bragg points), which are perpendicular to the ellipses. *III*, four rectangles (green) with rounded corners distributed around the centers of each quadrant. Features induced by inter-Brillouin Zone (BZ) scattering are plotted in dotted lines. They are weaker in the data compared to the pockets from intra-BZ scattering. (**e** to **g**) Real-space interference patterns corresponding to different FT pockets shown in the upper panels. Fourier filtering allows us to clearly visualize the prominent long range wave (**e**) and underlying atomic-scale wave (**g**), independently.

**Figure 3. Energy-dependent quantum interference patterns from surface quasi-particles of NbP. a**, A series of d$I$ / d$V$ maps measured at indicated voltages with tunneling current at 700pA (or 500pA when V = -40mV and 60mV). All images are of size 37 × 37 nm$^2$ and on a grid of 256 × 256 points. The two-dimensional electron gas on the NbP surface is scattered by the defects and form anisotropic standing waves (interference patterns in real space). The pattern shows a $C_2$ rather than $C_4$ symmetry. The wave lengths increase with elevated voltages, indicative of hole-like surface quasi-particles. The defects are shown as dark (bright) spots on the maps taken with negative (positive) voltages. **b**, The corresponding Fourier transform of each map in **a**. The voltage-dependent evolution of interference patterns (in momentum space) establishes a picture of the exotic surface states in this energy range. At -160mV, three patterns (ellipses, bow-ties and rectangles) are enlarged and merge into each other. Between -100mV and 140mV, all interference pockets are clearly resolved. Ellipses and rectangles fade away at 200meV while the bow-tie pockets remain visible.

**Figure 4. QPI pattern arising from scattering between different Fermi pockets. a**, Calculated joint density of states from the Fermi surface of NbP(001) surface. **b**, Same as **a** but with matrix-element effect considered. It thus resolves the spin and orbital dependent electron scattering in momentum space. **c**, Fourier transformed d$I$/d$V$ map measured close to Fermi level (-10mV). Bragg points are marked with their coordinates in $Q$-space. **d**, Calculated Fermi surface of NbP (001) with intra- and inter-pocket scatterings indicated by white arrows. $Q_1$ ($Q_2$) stands for inter-pocket scattering within the pocket surrounding $\bar{X}$ ($\bar{Y}$) while $Q_3$ is inter-pocket scattering between the bowtie-shaped and ellipse-shaped Fermi pockets. Red dotted arrows $S_1$-$S_3$ represent suppressed scattering processes involving the tadpole pockets. Grey stars mark the locations of the high symmetry points in first surface BZ. **e**, Calculated surface band structure of NbP along $\bar{X} - \bar{M} - \bar{Y}$. **f**, The dispersions of $Q_1$ (blue) and $Q_2$ (green) scattering vectors with respect to the bias voltage (energy), indicative of a shrinking hole pocket as energy increases. Experimental results (markers) are in good agreement with the theoretical calculations (solid lines).

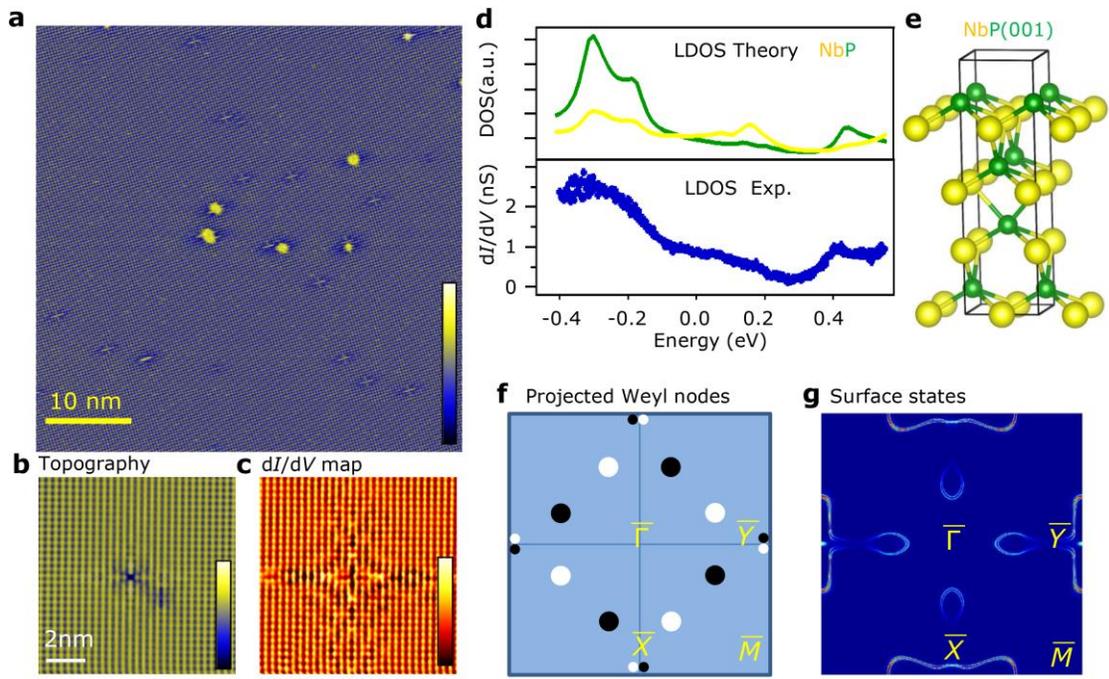

Figure 1

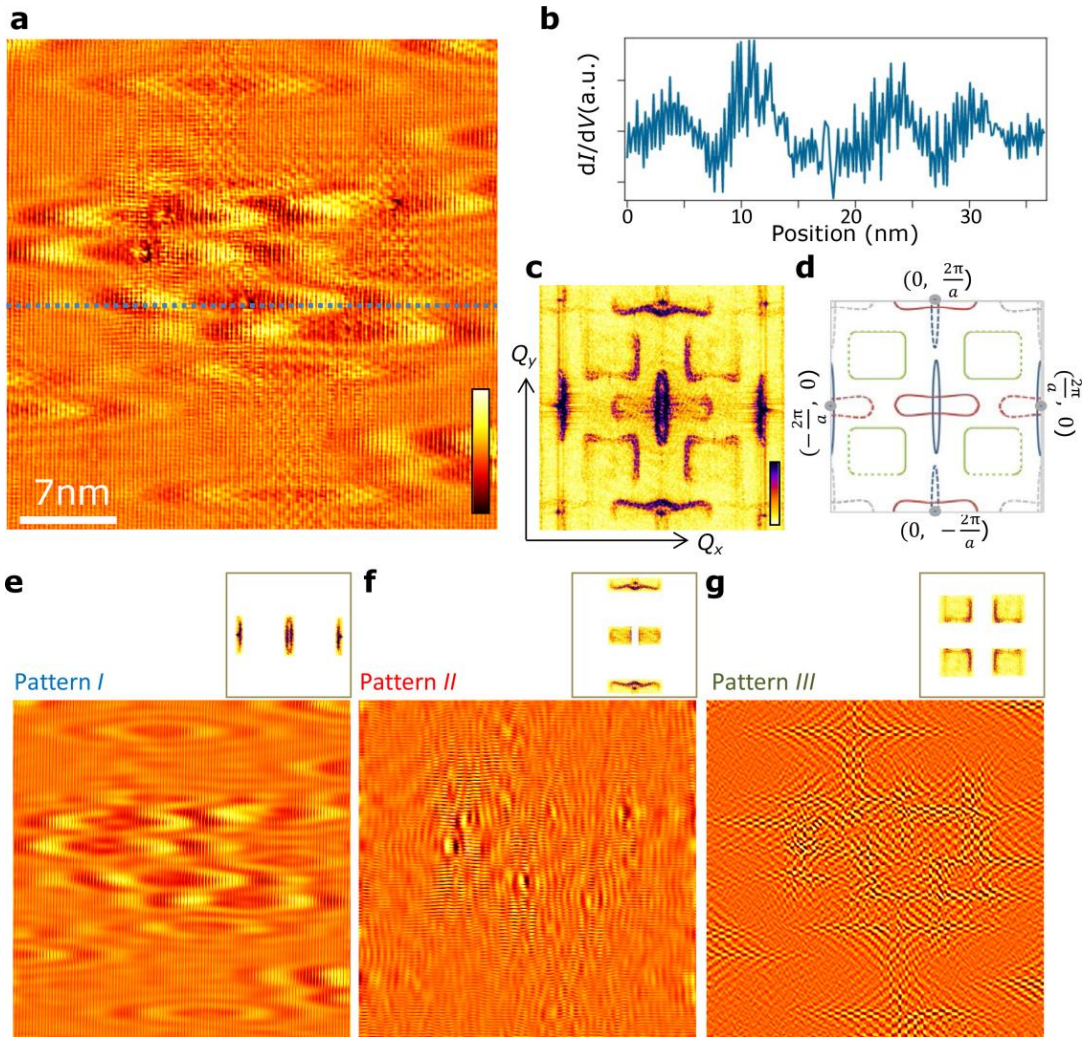

Figure 2

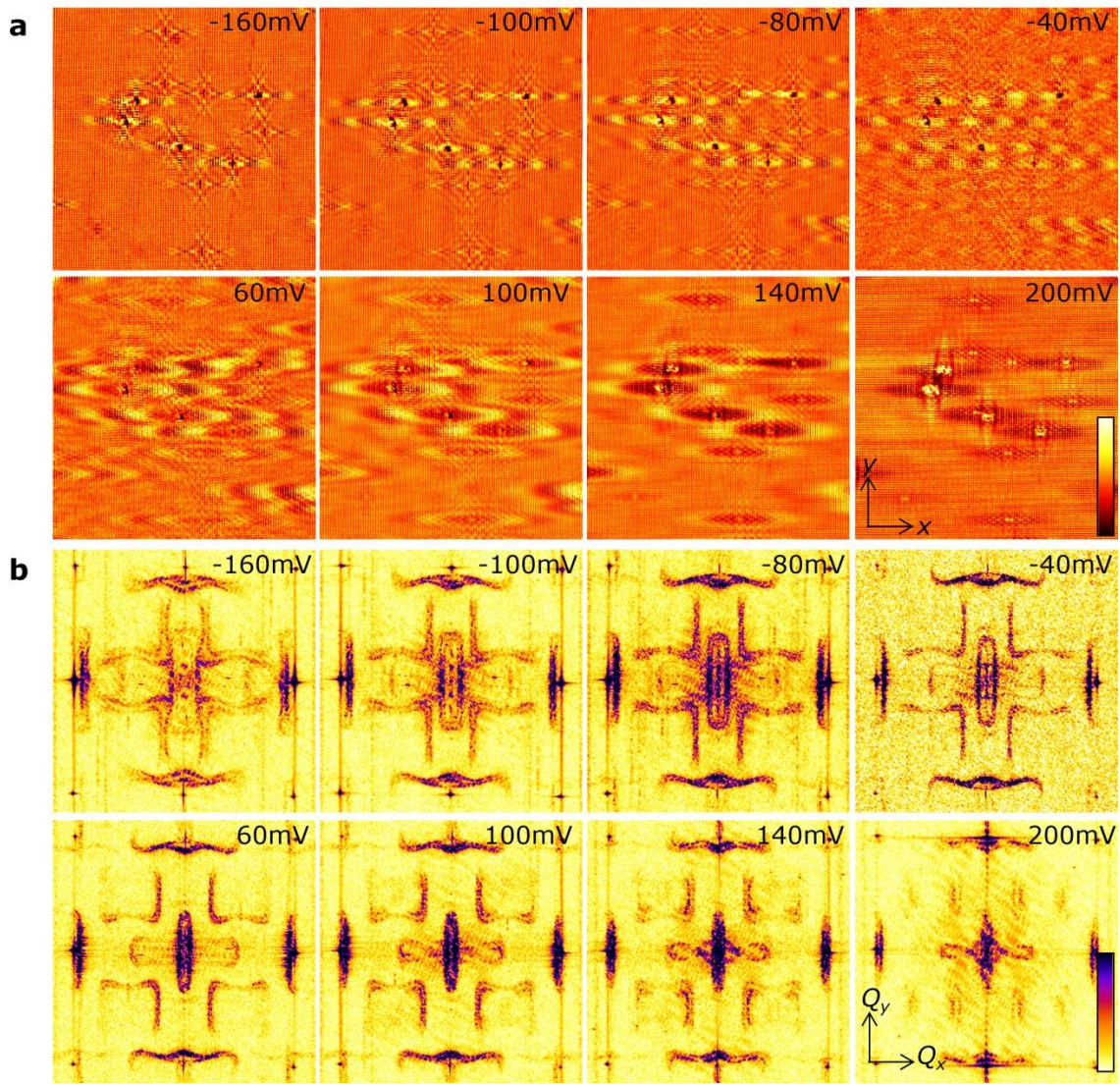

Figure 3

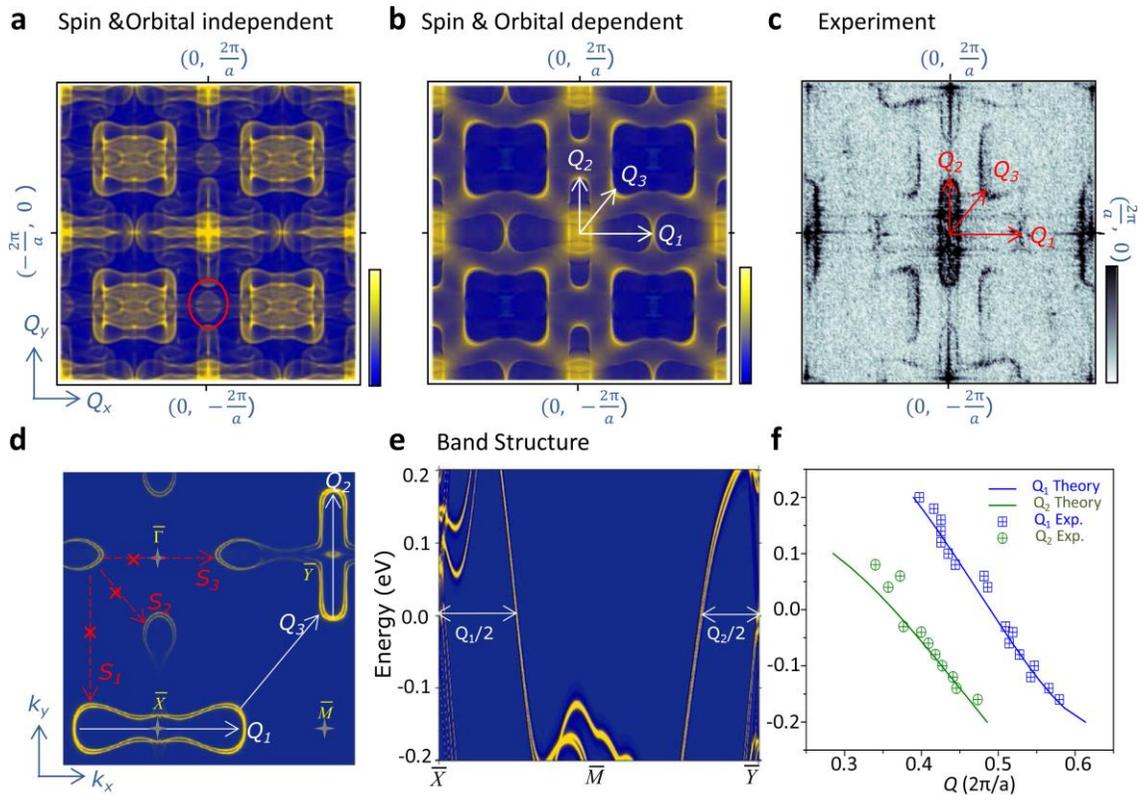

Figure 4